\documentclass[aps,prb,twocolumn,superscriptaddress]{revtex4}

\usepackage{graphicx}
\usepackage{amssymb}
\usepackage{amsmath}
\usepackage{enumerate}
\usepackage[dvipdfm,colorlinks,citecolor=blue,linkcolor=blue,urlcolor=blue]{hyperref}

\begin{document}

\title{Electronic structure of BaNi$_2$As$_2$}

\author{Bo Zhou}
\author{Min Xu}
\author{Yan Zhang}
\affiliation{State Key Laboratory of Surface Physics, Key Laboratory of Micro
and Nano Photonic Structures (MOE), and Department of Physics, Fudan
University, Shanghai 200433, People's Republic of China}

\author{Gang Xu}
\affiliation{Beijing National Laboratory for Condensed Matter Physics, Institute of Physics, Chinese Academy of Sciences, Beijing 100190, China}

\author{Cheng He}
\author{L. X. Yang}
\author{Fei Chen}
\author{B. P. Xie}
\affiliation{State Key Laboratory of Surface Physics, Key Laboratory of Micro
and Nano Photonic Structures (MOE), and Department of Physics, Fudan
University, Shanghai 200433, People's Republic of China}

\author{Xiao-Yu Cui}
\affiliation{Swiss Light Source, Paul-Scherrer Institut, 5232 Villigen, Switzerland}

\author{Masashi Arita}
\author{Kenya Shimada}
\author{Hirofumi Namatame}
\author{Masaki Taniguchi}
\affiliation{Hiroshima Synchrotron Radiation Center and Graduate School of Science, Hiroshima University, Hiroshima 739-8526, Japan}

\author{X. Dai}
\affiliation{Beijing National Laboratory for Condensed Matter Physics, Institute of Physics, Chinese Academy of Sciences, Beijing 100190, China}

\author{D. L. Feng}
\email[]{dlfeng@fudan.edu.cn}
\affiliation{State Key Laboratory of Surface Physics, Key Laboratory of Micro
and Nano Photonic Structures (MOE), and Department of Physics, Fudan
University, Shanghai 200433, People's Republic of China}

\date{\today}

\begin{abstract}

BaNi$_2$As$_2$, with  a first order phase transition around 131~K, is studied
by the angle-resolved photoemission spectroscopy. The measured electronic
structure is compared to the local density approximation calculations,
revealing similar large electronlike bands around $\bar{M}$ and differences along $\bar{\Gamma}$-$\bar{X}$. We further show that the electronic structure of BaNi$_2$As$_2$ is distinct from that of the sibling iron pnictides.
Particularly, there is no signature of band folding, indicating no collinear
SDW related magnetic ordering. Moreover, across the strong first order phase
transition, the band shift exhibits a hysteresis, which is directly related to
the significant lattice distortion in BaNi$_2$As$_2$.

\end{abstract}


\maketitle

\section{Introduction}

The  discovery of iron-based high-temperature superconductors has ignited
intensive
studies.\cite{Kamihara,Chen_Sm1,ZARen_Sm,Chen_Sm2,MXu,CLWang_Sm,ZARen_Nd,Kursumovic_Nd,ZARen_Pr}
The superconducting transition temperature ($T_c$) has risen up to 56~K in
$R$FeAsO$_{1-x}$F$_x$ ($R$=Sm, Nd, Pr, ...).\cite{ZARen_Sm,Chen_Sm2,CLWang_Sm,ZARen_Nd,Kursumovic_Nd,ZARen_Pr} On the
other hand, the sibling Ni-based compounds only possess relatively low $T_c$,
i.e., LaONiP ($T_c$=3~K),\cite{LaONiP} LaONiAs ($T_c$=2.75~K),\cite{LaONiAs}
BaNi$_2$P$_2$ ($T_c$=2.4~K),\cite{BaNi2P2} BaNi$_2$As$_2$ ($T_c$=0.7~K),\cite{Ronning} and SrNi$_2$As$_2$ ($T_c$=0.62~K).\cite{SrNi2As2}. It is intriguing to understand why $T_c$ is low in the Ni-based compounds,
which may facilitate understanding the high-$T_c$ in iron-based ones.

BaFe$_2$As$_2$, a typical parent compound of iron-based superconductor with the ThCr$_2$Si$_2$ structure, exhibits a structural transition from tetragonal to
orthorhombic, concomitant with a spin-density-wave (SDW) transition at
140~K.\cite{Rotter} The structural transition was suggested to be driven by the magnetic degree of freedom.\cite{Yildirim,CFang,Cenke,CHe,LX1111,NLWang_1111}
BaNi$_2$As$_2$ with the same structure, displays a  structural transition
around 131~K, however, from a tetragonal phase to a lower symmetry triclinic
phase.\cite{Sefat} No evidence of SDW is reported in BaNi$_2$As$_2$ so far. On the other hand, it was pointed out that the transition in
BaNi$_2$As$_2$ is a first-order one, while that is more second-order-like in
BaFe$_2$As$_2$.\cite{Sefat} Moreover, the c-axis resistivity drops by two
orders of magnitude and a thermal hysteresis of in-plane resistivity is present at the transition of BaNi$_2$As$_2$.\cite{Ronning,Sefat} Although there are
intriguing resemblance as well as differences between BaFe$_2$As$_2$ and
BaNi$_2$As$_2$, the electronic structure of the latter is still not exposed.

Here we report the angle-resolved photoemission spectroscopy (ARPES) study of
the electronic structure of BaNi$_2$As$_2$. Our data are compared to the band
structure calculation of BaNi$_2$As$_2$ and the results of iron pnictides
reported before, revealing their similarities and differences. Particularly,
no band folding is found in the electronic structure of BaNi$_2$As$_2$,
confirming that there is no collinear SDW type of magnetic ordering. Because of the intimate relation between superconductivity and magnetism,\cite{Drew_Nat,Christianson,LiFeAs_NMR,Graser} the absence of
magnetic ordering is possibly related to the low-$T_c$ of BaNi$_2$As$_2$.
Furthermore, a hysteresis is observed for the band shift, resembling the
hysteresis in the resistivity data. The band shift can be accounted for by the
significant lattice distortion in BaNi$_2$As$_2$, in contrast to iron
pnictides, where the band shift is largely caused by the magnetic ordering.

\section{Experimental}

\begin{figure}[b]
\includegraphics[width=2.5cm]{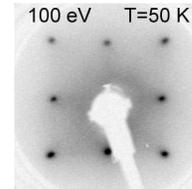}
\caption{The low-energy electron diffraction pattern of BaNi$_2$As$_2$ in the triclinic phase taken with 100~eV incident electrons. }\label{LEED}
\end{figure}

\begin{figure*}
\includegraphics[width=17cm]{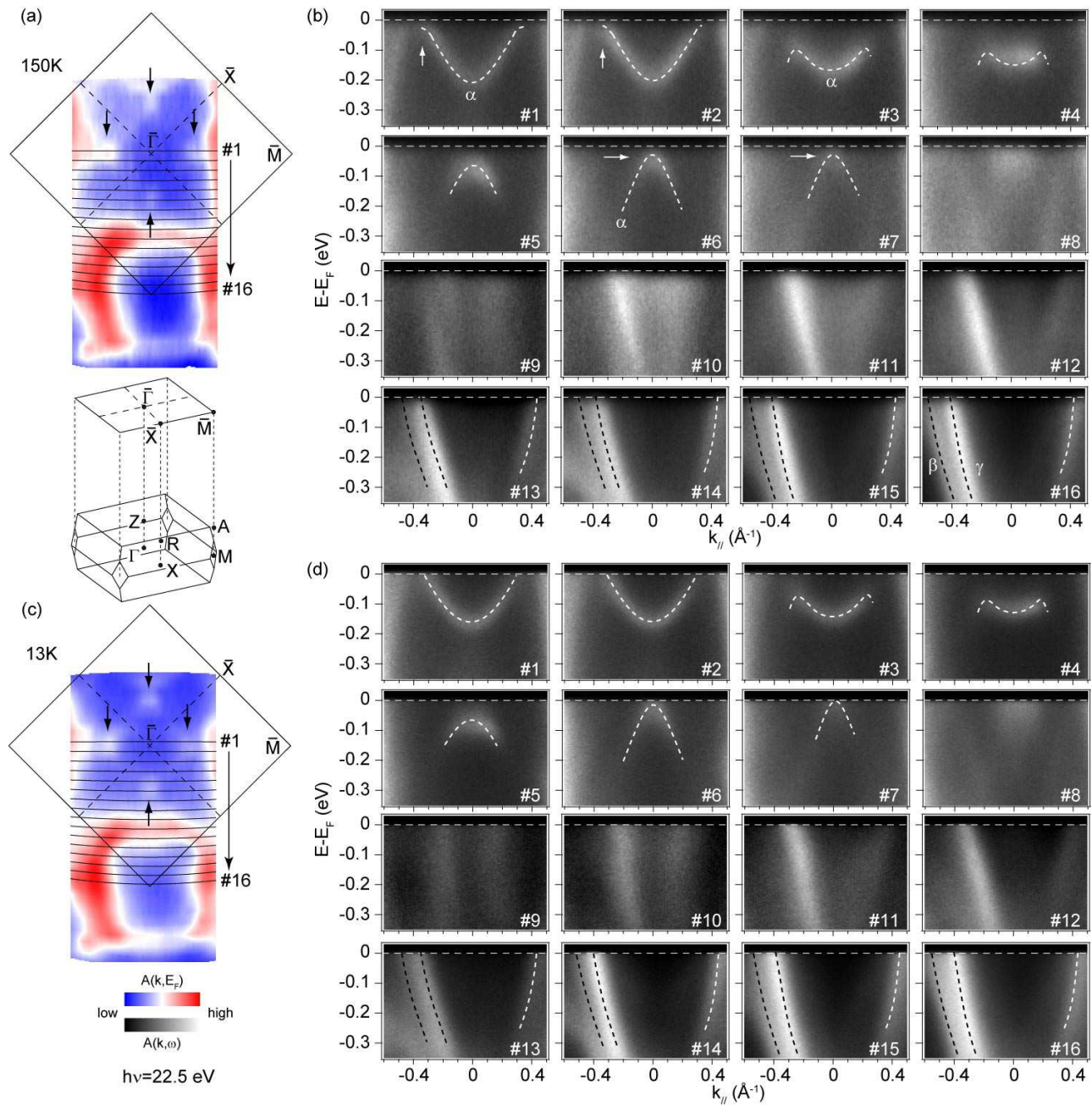}
\caption{(Color online) Photoemission data of BaNi$_2$As$_2$. (a) Photoemission
intensity map at the Fermi energy ($E_F$) measured at 150~K in the tetragonal
phase. For the convenience of data presentation, the two-dimensional (2D)
Brillouin zone is referred to hereafter, which is the projection of
three-dimensional (3D) Brillouin zone. (b) Photoemission intensity plots of
cuts 1-16 as indicated in panel a. (c) and (d) are the same as panels a and b
but for data measured at 13~K in the triclinic phase. Band dispersions are
indicated by the dashed curves. Data were taken in HSRC with circularly polarized 22.5~eV photons. Labels are explained in the text.}\label{GM}
\end{figure*}

BaNi$_2$As$_2$ single crystals were synthesized by self-flux method, and a
similar synthesis procedure has been described in Ref.~\onlinecite{Sefat}. Its
stoichiometry was confirmed by energy dispersive x-ray (EDX) analysis. ARPES
measurements were performed (1) with circularly-polarized synchrotron light and
randomly-polarized 8.4~eV photons from a xenon discharge lamp at Beamline 9 of
Hiroshima synchrotron radiation center (HSRC), (2) with linearly polarized
synchrotron light at the surface and interface spectroscopy (SIS) beamline of
Swiss light source (SLS), and (3) with randomly-polarized 21.2~eV photons from
a helium discharge lamp. Scienta R4000 electron analyzers are equipped in all
setups. The typical energy resolution is 15~meV, and angular resolution is
0.3$^\circ$. The samples were cleaved \textit{in situ}, and measured in
ultrahigh vacuum better than 5$\times10^{-11}$~mbar. The high quality sample
surface was confirmed by the clear pattern of low-energy electron diffraction
(LEED), where no sign of surface reconstruction is observed (Fig.~\ref{LEED}).

\section{Experimental results}

\begin{figure}
\includegraphics[width=8.5cm]{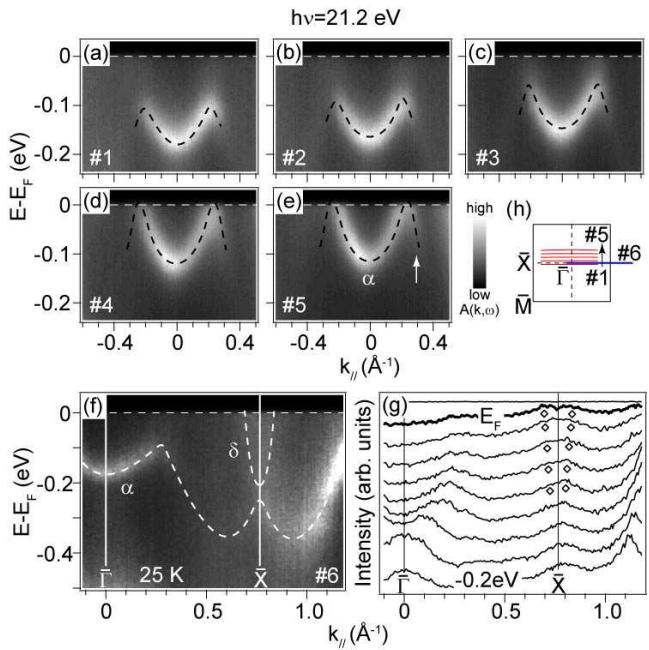}
\caption{(Color online) Photoemission data along several cuts parallel to $\bar{\Gamma}$-$\bar{X}$ in the triclinic phase.
(a)-(e) Photoemission intensity plots of cuts 1-5 as indicated in panel g, taken in HSRC with  circularly polarized 21.2~eV photons at 20~K.
(f) Photoemission intensity plot of cut 6 as indicated in panel g, taken with randomly polarized 21.2~eV photons from a helium lamp at 25~K. (g) The MDCs corresponding to panel f. (h) Cuts 1-6 are indicated in the projected 2D Brillouin zone. The dashed curves and markers trace the band dispersions.}\label{GX}
\end{figure}

\begin{figure}
\includegraphics[width=8.5cm]{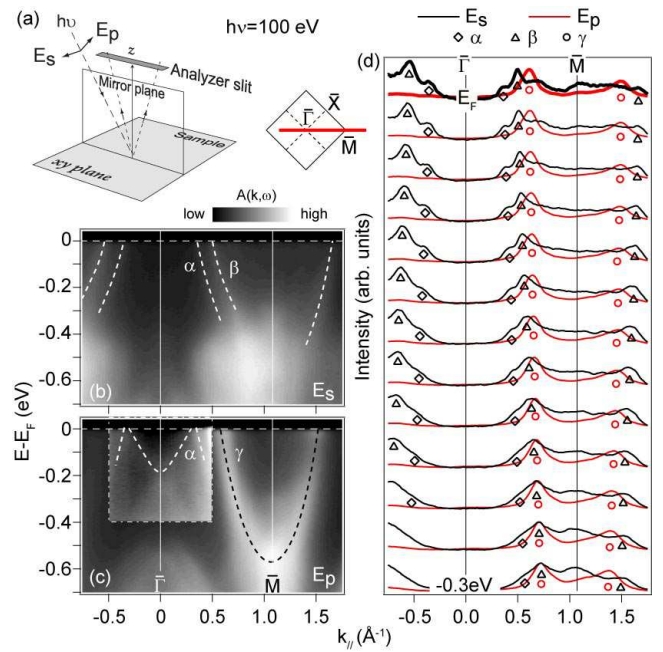}
\caption{(Color online) Photoemission data along $\bar{\Gamma}$-$\bar{M}$ measured at 10~K in the triclinic phase with 100~eV linearly polarized light at SLS. (a) Experimental setup for the s and p polarization geometries, and the indication of the $\bar{\Gamma}$-$\bar{M}$ cut in the projected 2D Brillouin zone. (b) and (c) Photoemission intensity plots measured in the s and p polarization geometries respectively. The image contrast in the rectangular region as enclosed by dash-dotted lines in panel c is adjusted to reveal the bands in this region. (d) Stack of MDCs in the s and p polarization geometries. Each MDC is normalized by its integrated weight. Dashed curves and markers trace the band dispersions.  Labels are explained in the text.}\label{aroundM}
\end{figure}

Figures~\ref{GM}(a) and \ref{GM}(c) show the Fermi surface maps measured with
circularly polarized 22.5~eV photons in the tetragonal and triclinic phases,
respectively. There are four patches near the $\bar{\Gamma}$ point as indicated by the arrows, and two electron pockets around the $\bar{M}$ point. Sixteen cuts from $\bar{\Gamma}$ to $\bar{M}$ are presented in Fig.~\ref{GM}(b) to illustrate the electronic structure evolution in the tetragonal phase, and the corresponding data in the triclinic phase are shown in Fig.~\ref{GM}(d). In the tetragonal phase, the parabolic-shaped band in cuts 1-2 around $\bar{\Gamma}$ is referred as $\alpha$. The $\alpha$ band appears to be $M$-shaped in cuts 3-4. From cut 1 to cut 4, the parabolic part shrinks continuously, and eventually only the inverted parabolic part is observable in cuts 5-7. The evolution of the electronlike bands around $\bar{M}$ are shown in cuts 8-16, many of which are complex due to the rapid change of dispersions. Nonetheless, two bands ($\beta$ and $\gamma$) can be resolved as indicated by the dashed curves in cut 16 and will be further elaborated in Fig.~\ref{aroundM}. The data in the triclinic phase [Fig.~\ref{GM}(d)] are generally similar to those in the tetragonal phase. Nonetheless, we note that $\alpha$ already shows some bending near the Fermi energy ($E_F$) in the tetragonal phase as indicated by the arrows on the data taken along cuts 1-2. On the other hand, $\alpha$ just passes through $E_F$ without bending in the triclinic phase. Moreover, the band top of $\alpha$ is below $E_F$ in cuts 6-7 of Fig.~\ref{GM}(b) as indicated by the arrows, but barely touches $E_F$ in Fig.~\ref{GM}(d), where cuts 6-7 pass through one of the four patches around $\bar{\Gamma}$. Therefore, the four spectral weight patches around $\bar{\Gamma}$ [as marked by four arrows in Fig.~\ref{GM}(a)] are due to the residual spectral weight of the $\alpha$ band in the tetragonal phase. However, the $\alpha$ band shifts up and they evolve into small holelike Fermi surfaces in the triclinic phase. This will be further illustrated in Fig.~\ref{GX}. Of note, from this complete set of data, we do not observe any sign of band folding or splitting like that in the iron pnictides.\cite{LXYang,YZhang,BZhou,GDLiu,ZX_SDW}

To further illustrate the electronic structure of BaNi$_2$As$_2$, Figs.~\ref{GX}(a)-\ref{GX}(e) present photoemission intensities along five cuts parallel to the $\bar{\Gamma}$-$\bar{X}$ direction in the triclinic phase. The $M$-shaped feature originated from $\alpha$ simply moves towards $E_F$ from cut 1 to cut 5, touching $E_F$ at cuts 4-5, which confirms that the four small Fermi surfaces around $\bar{\Gamma}$ are holelike. Note that the downward part of $\alpha$ [indicated by the arrow in Fig.~\ref{GX}(e)] is clearly resolved here, while it is barely observable in the same momentum region when the cuts are along the $\bar{\Gamma}$-$\bar{M}$ direction, as shown in cuts 1-2 of Figs.~\ref{GM}(b) and \ref{GM}(d). It highlights the matrix element effects since the 3$d$ orbitals have specific orientations. Figure~\ref{GX}(f) shows the photoemission intensity along $\bar{\Gamma}$-$\bar{X}$, taken with randomly polarized 21.2~eV photons from a helium lamp in the triclinic phase. The determined band structure is traced by dashed curves, where the broad spectral weight around $\bar{X}$ are attributed to an electronlike band $\delta$, which is further shown by markers in the corresponding momentum distribution curves (MDCs) [Fig.~\ref{GX}(g)]. Therefore, there is an electron Fermi pocket around $\bar{X}$.

Similar to iron pnictides, the bands near $E_F$ are quite complicated and
mainly contributed to by the Ni 3$d$ electrons in BaNi$_2$As$_2$. To resolve the complex bands around $\bar{M}$, we utilize the linearly polarized light, which could only detect bands with certain symmetry, so that the measured partial electronic structure helps reducing the complexity in analysis.\cite{YZhang_orbital} Figure~\ref{aroundM} presents data along $\bar{\Gamma}$-$\bar{M}$, taken with linearly polarized 100~eV photons in SLS in the triclinic phase. Two polarization geometries (s and p) are illustrated in Fig.~\ref{aroundM}(a). In the s polarization geometry, we resolve two bands, whose dispersions are depicted by dashed curves in the photoemission intensity plot [Fig.~\ref{aroundM}(b)]. While in the p polarization geometry, one intense parabolic electronlike band around $\bar{M}$ is resolved with the band bottom at about -0.57~eV. The dispersions in both geometries are marked in the corresponding MDCs [Fig.~\ref{aroundM}(d)]. The asymmetry of the dispersion indicated by triangles may be due to the slight sample misalignment. The image contrast in the dash-dotted region in Fig.~\ref{aroundM}(c) is adjusted to highlight the $\alpha$ feature. The observed $\alpha$ feature is consistent with the data in Figs.~\ref{GM} and \ref{GX}. By comparing with the Fermi crossings observed in cuts 1 and 16 of Fig.~\ref{GM}(d), we attribute the three bands to $\alpha$, $\beta$, and $\gamma$ as shown in Fig.~\ref{aroundM}(d), where $\beta$ and $\gamma$ are two electronlike bands around $\bar{M}$. Moreover, since the experimental setup under the s (p) polarization geometry detects states with odd (even) symmetry with respect to the mirror plane, the $\beta$  band is of mainly odd symmetry while $\gamma$ is of even symmetry. We note that $\alpha$  is observed in both geometries [Fig.~\ref{aroundM}(b) and \ref{aroundM}(c)], suggesting that the $\alpha$ band has mixed symmetries.

\begin{figure}
\includegraphics[width=8.5cm]{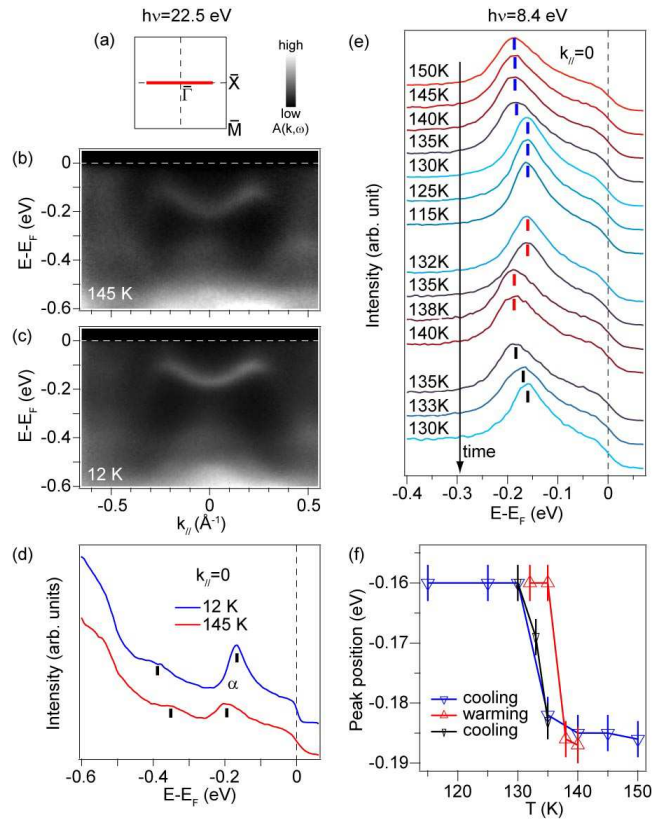}
\caption{(Color online) Temperature dependence of photoemission data.
(a) Indication of the cut by the thick line in the projected 2D Brillouin zone.
(b) and (c) Photoemission intensity plots along the cut shown in panel a measured at 145 and 12~K respectively.
(d) Comparison of EDCs at $k_\parallel$=0 for data shown in panels b and c.
Data were taken in HSRC with circularly polarized 22.5~eV photons for panels b-d.
(e) EDCs at $k_\parallel$=0 for a cooling-warming-cooling cycle, measured with randomly polarized 8.4~eV photons from a xenon discharge lamp.
The short bars indicate the peak positions. (f) Summary of peak positions obtained from panel e, which exhibits a hysteresis.}\label{loop}
\end{figure}

\begin{figure*}
\includegraphics[width=17cm]{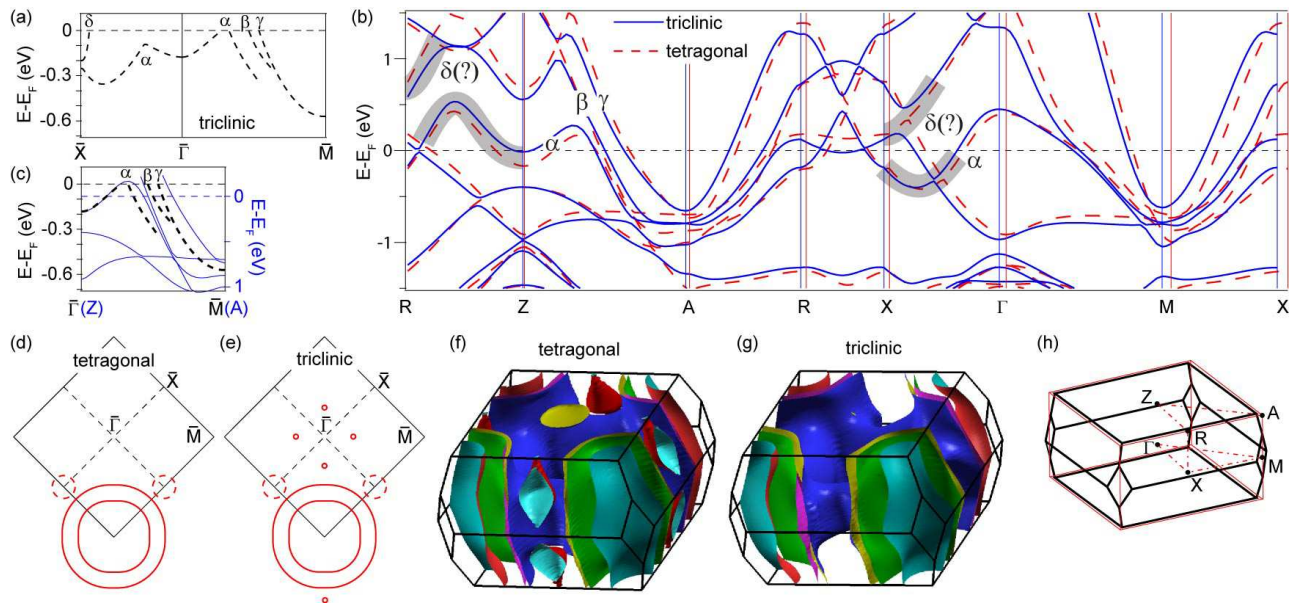}
\caption{(Color online) (a)  Band structure along
$\bar{X}$-$\bar{\Gamma}$-$\bar{M}$ in the triclinic phase summarized from
Figs.~\ref{GX} and \ref{aroundM}. (b) Calculated band structures in the
tetragonal (dashed curves) and triclinic (solid curves) phases.\cite{optical}
(c) Experimental band structure from panel a (dashed curves) overlaid by the
calculated band structure (solid curves) in the triclinic phase. The energy
scale of the calculated band structure at the right side is 1.66 times of the
energy scale of the experimental one on the left side.  The summary of
experimental Fermi contours (d) in the tetragonal phase, and (e) in the
triclinic phase respectively. The small circles indicate the small Fermi
surfaces around $\bar{\Gamma}$ and the dashed circles indicate the Fermi
surface around $\bar{X}$. The sketches are depicted in part of the projected 2D
Brillouin zone. (f) and (g) The calculated 3D Fermi surfaces in the tetragonal
and triclinic phases respectively.\cite{optical} (h) The 3D tetragonal
Brillouin zone (thick lines) with notations for high symmetry points. Labels
are explained in the text.}\label{calc}
\end{figure*}

\begin{figure}
\includegraphics[width=8.5cm]{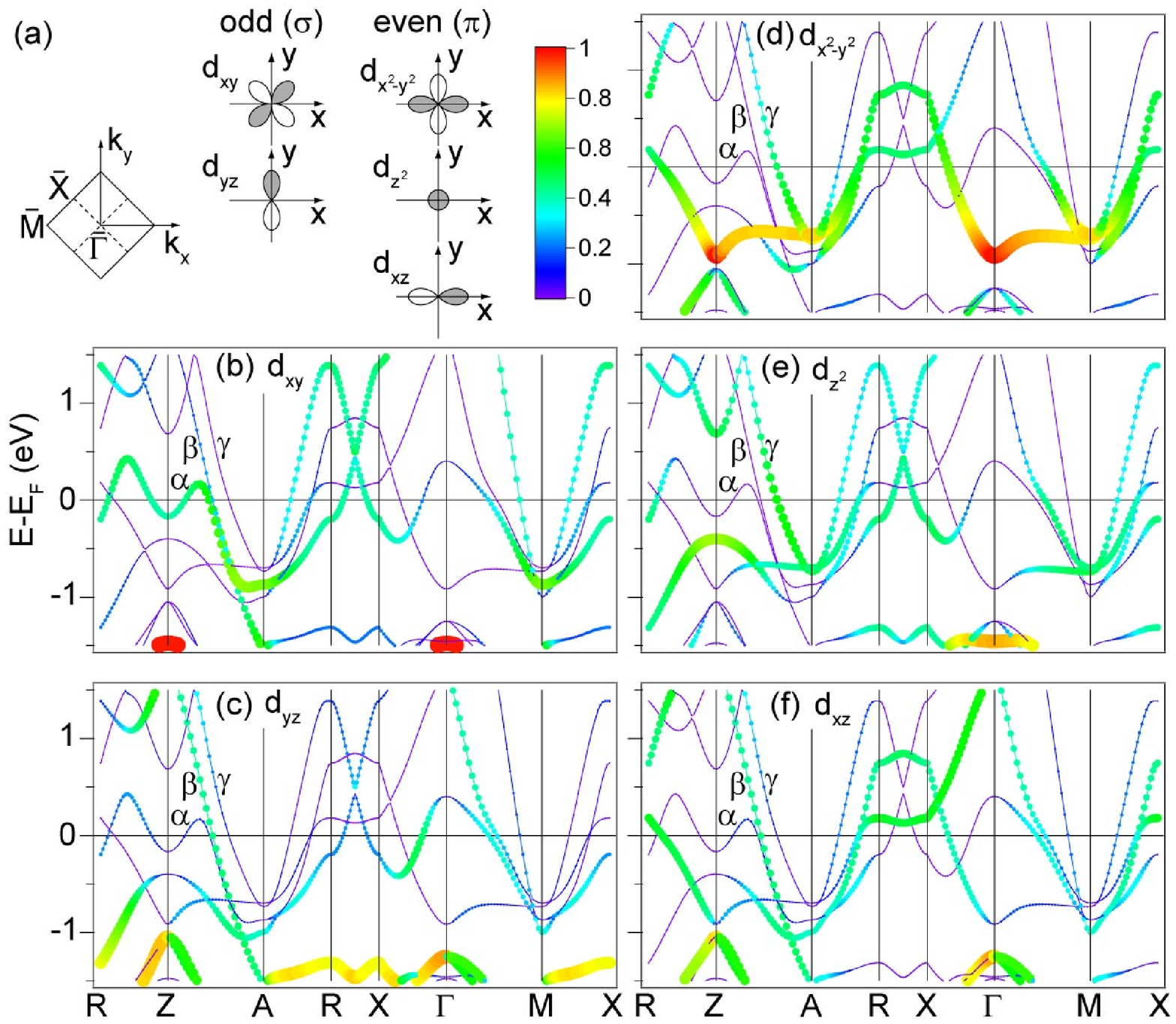}
\caption{(Color online) (a) Illustration of the in-plane spatial symmetry of 3$d$ orbitals.
(b)-(f) Contributions of the $d_{xy}$, $d_{yz}$, $d_{x^2-y^2}$, $d_{z^2}$, and $d_{xz}$ orbitals
to the calculated band structure of BaNi$_2$As$_2$ respectively. The contribution is represented
by both the size of the symbols and the color scale.}\label{orbital}
\end{figure}

To study the first order transition of BaNi$_2$As$_2$, the temperature
dependence is presented in Fig.~\ref{loop}. The photoemission intensity plots
along $\bar{\Gamma}$-$\bar{X}$ are shown in Figs.~\ref{loop}(b) and
\ref{loop}(c) for the tetragonal and the triclinic  phases respectively. The
corresponding energy distribution curves (EDCs) at $k_\parallel$=0 are stacked
in Fig.~\ref{loop}(d). Interestingly, the band bottom of the $M$-shaped feature is moved from -200~meV in the tetragonal phase to -170~meV in the triclinic
phase. In other words, the $M$-shaped band moves towards $E_F$ and its
electronic energy is raised up. However, another feature at higher binding
energies shifts away from $E_F$. Its band top is moved from -350~meV at 145~K
to -390~meV at 12~K, which partially saves the electronic energy. Since the
resistivity shows a hysteresis loop,\cite{Ronning} it is intriguing to
investigate whether a  similar hysteresis could be observed for the electronic
structure. Data in Figs.~\ref{loop}(e)-\ref{loop}(f) are taken with randomly polarized 8.4~eV photons from a xenon discharge lamp, in a cooling-warming-cooling cycle. The EDCs at $k_\parallel$=0 across the transition  are stacked in Fig.~\ref{loop}(e), where the peak positions are indicated by  short bars. The temperature dependence of peak positions is summarized in Fig.~\ref{loop}(f), showing a clear hysteresis with the band shift as much as 25~meV. Such electronic structure demonstration of a hysteresis of 3~K is so far the most obvious. A hysteresis in the electronic structure has been observed in FeTe, but with a loop width of only 0.5~K.\cite{YZhang_FeTe} Our observation here is consistent with the bulk transport properties, which indicates that the measured electronic structure reflects the bulk properties.

\section{Band structure calculations}

The measured band structure and Fermi surface are summarized in
Figs.~\ref{calc}(a), \ref{calc}(d) and \ref{calc}(e). For comparison, local
density approximation calculations which have been reported before in
Ref.~\onlinecite{optical} are reproduced in Figs.~\ref{calc}(b), \ref{calc}(f)
and \ref{calc}(g).  The notations for bands near $E_F$ are labeled in
Figs.~\ref{calc}(a) and \ref{calc}(b). Qualitatively, although not all
calculated bands were observed, the main features of the experiments are
captured by the calculation, such as the dispersion nature of the bands. The
$\alpha$, $\beta$, and $\gamma$ bands of the experimental results in the
$\bar{\Gamma}$-$\bar{M}$ direction are similar to the numerical results in the
$Z$-$A$ direction. As shown in Fig.~\ref{calc}(c), the measured $\alpha$ band
along $\bar{\Gamma}$-$\bar{M}$ matches the calculation well after the
calculated bands are renormalized by a factor of 1.66 and shifted down by
0.08~eV. This renormalization factor is consistent with the results of optical
measurements.\cite{optical} Although not all bands could match, it may suggest
that the correlation in BaNi$_2$As$_2$ is weaker than that in iron
pnictides.\cite{ZX_SDW,FChen_FeTeSe} Along $\bar{\Gamma}$-$\bar{X}$, the
observed $\alpha$ and $\delta$ bands partially resemble the calculated
dispersions  along both $Z$-$R$ and $\Gamma$-$X$, as highlighted by  the shaded
regions [Fig.~\ref{calc}(b)], but the energy positions do not match. Our data
along in this direction might correspond to a $k_z$ between $\Gamma$ and $Z$.

As expected from the differences in the experimentally determined and
calculated band structures, the Fermi surface topologies are quite different in
both the experiments and the calculations. In our data [Figs.~\ref{calc}(d) and
\ref{calc}(e)], we observe four small Fermi pockets around $\bar{\Gamma}$ only
in the triclinic phase. Around $\bar{M}$, two electronlike Fermi pockets are
resolved in both phases. Around $\bar{X}$, the observed Fermi crossings are
from an electronlike pocket. As a comparison, in the calculated Fermi surface
of high-T tetragonal phase [Figs.~\ref{calc}(f)], there are two large warped
cylinders of electron pockets around the zone corner, a pocket interconnected
from the zone center to a large deformed cylinder around the zone corner, a 3D
electron pocket around $Z$, and 3D pockets located between $X$ and $R$. In the
low-T triclinic phase [Figs.~\ref{calc}(g)], 3D pockets around $Z$ and between
$X$ and $R$ are gapped out. The large electron Fermi pockets around $\bar{M}$
observed in our data are generally consistent with that in the calculation,
which is a direct consequence of two more electrons from Ni than Fe. Note that
the $k_z$-dispersion is significant in the calculation. However, we have
measured with four different photon energies, including the more bulk-sensitive
8.4~eV photons, and no obvious differences in dispersion have been observed.
Therefore, the $k_z$-dispersions in BaNi$_2$As$_2$ may be weaker than
calculated.

For a multiband and multiorbital superconductor, it is crucial to understand
the orbital characters of the band structure.  Because of the symmetry of 3$d$
orbitals with respect to the mirror plane, the s polarization geometry in
photoemission can only detect the $d_{xy}$ and $d_{yz}$ orbitals while the p
polarization geometry can only detect the $d_{x^2-y^2}$, $d_{z^2}$, and
$d_{xz}$ orbitals [Fig.~\ref{orbital}(a)].\cite{YZhang_orbital} The
contributions of the five 3$d$ orbitals to the calculated band structure are
presented in Fig.~\ref{orbital}(b)-\ref{orbital}(f), which therefore can be
compared to our polarization dependent data. Along $Z$-$A$, the $\alpha$ band
is consisted of mainly the odd $d_{xy}$ orbital and some contributions of odd
$d_{yz}$ and even $d_{xz}$, thus can be observed in both the s and p
polarization geometries; while the $\gamma$ band is consisted of the even
$d_{x^2-y^2}$ and $d_{z^2}$ orbitals, thus can only be observed in the p
polarization geometry. They are in good agreement with our observation. The
$\beta$ band is consisted of the odd $d_{xy}$, $d_{yz}$ and even $d_{xz}$
orbitals in the calculation, thus should be observed in both the s and p
polarization geometries. However, $\beta$ is mainly detected in the s
polarization geometry, possibly because in the p polarization geometry it is
buried in the intense peak of $\gamma$. The consistency between our data and
the calculated orbital characters confirms that our data along
$\bar{\Gamma}$-$\bar{M}$ match the band structure calculation along $Z$-$A$.

\section{Discussions}

It was observed in the optical data that the phase transition leads to a
reduction of conducting carriers, consistent with the removal of small Fermi
surfaces shown by the calculation.\cite{optical} However, we do not observe
such behavior by ARPES. On the contrary, instead of the disappearance of small
Fermi surfaces in the triclinic phase, we observe that bands shift up in
energy, leading to additional four  Fermi surfaces. The inconsistency between the optical data and our photoemission data suggests that the changes in optical data across the phase transition are possibly an integrated effect of band structure reorganization over the entire Brillouin zone, instead of the disappearance of certain Fermi surface sheets; but it is also possible that only limited $k$-space has been probed in the current photoemission study.

As a sibling  compound of iron pnictides, BaNi$_2$As$_2$ exhibits quite
different properties and electronic structure. The parent compounds of iron
pnictides show a second-order-like transition that is the SDW transition
concomitant with a structural transition. However, BaNi$_2$As$_2$ shows a
strong first-order-like structural transition, without magnetic ordering
reported to date. From the aspect of electronic structure, iron pnictides
possess several hole pockets around $\bar{\Gamma}$, and several electron
pockets around $\bar{M}$, but have no pockets around $\bar{X}$, while the band
structure of BaNi$_2$As$_2$ is dramatically different from that of the iron
pnictides. Moreover, no signature of folding could be found in our data,
confirming that no collinear magnetic ordering exists in BaNi$_2$As$_2$.
Because of the intimate relation between the magnetism and superconductivity,\cite{Drew_Nat,Christianson,LiFeAs_NMR,Graser}
the absence of magnetic ordering might be related to the low-$T_c$ in
BaNi$_2$As$_2$.

Across the structural transition in  BaNi$_2$As$_2$, the  Ni-Ni distance
changes from 2.93~\AA$^{-1}$ to 2.8~\AA$^{-1}$ (or 3.1~\AA$^{-1}$),
corresponding to a lattice distortion as much as $\sim$5\% in
average.\cite{Sefat} A rough estimation can be made for the hopping parameter
$t_{dd}$ between certain $d$-$d$ orbitals after the lattice distortion
according to Ref.~\onlinecite{Harrison},
\begin{eqnarray*}
t_{dd}&=&t_{dd}^0(1+\delta)^{-\frac{7}{2}}\approx t_{dd}^0(1-\frac{7}{2}\delta), \\
\Longrightarrow \Delta t&=&t_{dd}-t_{dd}^0=-\frac{7}{2}\delta t_{dd}^0,
\end{eqnarray*}
where $\delta$ is the relative  lattice distortion; $t_{dd}^0$ is the hopping
parameter before the distortion; $\Delta t$ is the induced hopping parameter
change. Therefore, the 5\% lattice distortion would cause $\sim$17.5\% of
change to $t_{dd}^0$. Since the measured bandwidth of $\alpha$ is at least
200~meV, the induced band shift would be larger than 35~meV, more than enough
to account for the measured band shift of 25~meV. Note that the differences
between the calculated band structures of tetragonal and triclinic phases in
Fig.~\ref{calc}(f) are solely induced by considering the different lattice
parameters. For instance, the band along $Z$-$R$ near $E_F$ has a shift of 25\%
of  the bandwidth, generally consistent with our observation. Therefore, our
BaNi$_2$As$_2$ data provide a prototypical experimental showcase of band shift
due to significant lattice distortion. As a comparison, the lattice distortion
in NaFeAs is 0.36\%, which would induce only 1~meV of band shift, much smaller
than the observed 16~meV by ARPES.\cite{CHe} Similar results can be found in
other iron pnictides.\cite{LXYang,YZhang,BZhou,ZX_SDW,GDLiu}  The minor lattice
distortion cannot account for the large band shift observed in iron pnictides,
therefore it has been concluded that the only promising explanation  left is
that the band shift is related to the magnetism.\cite{CHe,BZhou_SNS}

\section{Conclusions}

To summarize, we report the first electronic structure study of BaNi$_2$As$_2$
by ARPES.  In comparison with the band calculation of BaNi$_2$As$_2$ and
reports of iron pnictides, we conclude several points as following:
\begin{enumerate}[i)~]
\item We observe four small Fermi pockets around $\bar{\Gamma}$ only in the triclinic phase, an electronlike pocket around $\bar{X}$ and two
    electronlike pockets around $\bar{M}$ in both tetragonal and triclinic
    phases. The main features of the measured band structure along $\bar{\Gamma}$-$\bar{M}$ is qualitatively captured by the band calculations, however differences exist along $\bar{\Gamma}$-$\bar{X}$. The electronic structure of BaNi$_2$As$_2$ is also distinct from the that of iron pnictides. Moreover, the correlation effects in BaNi$_2$As$_2$ seems to be weaker than that in iron pnictides, as the band renormalization factor is smaller for BaNi$_2$As$_2$.
\item Unlike  iron pnictides, we do not observe any sign of band folding in BaNi$_2$As$_2$, confirming no collinear SDW related magnetic ordering.
    Since the magnetism intimately relates to the superconductivity,
    possibly this is why the $T_c$ is much lower in BaNi$_2$As$_2$ than in
    iron pnictides.
\item The  SDW/structural transition in iron pnictides is
    second-order-like, while the structural transition in BaNi$_2$As$_2$ is first-order  and a thermal hysteresis is observed for its band shift.
\item The  band shift in BaNi$_2$As$_2$ is caused by the significant
    lattice distortion. On the other hand, the band shifts in the iron
    pnictides cannot be accounted for by the minor lattice distortion
    there, but are related to the magnetic ordering.
\end{enumerate}

\begin{acknowledgments}

Part of this work was performed at the Surface and Interface Spectroscopy beamline, Swiss Light Source, Paul Scherrer Institute, Villigen, Switzerland. We thank C. Hess and F. Dubi for technical support. This work was supported by the NSFC, MOE, MOST (National Basic Research Program No. 2006CB921300 and 2006CB601002), STCSM of China.

\end{acknowledgments}

\end{document}